# ANALYTICAL MODELLING OF THE IMPACT OF FOLIAGE COVER ON THE PROPAGATION LOSS IN A SMART FARMING WIRELESS IOT APPLICATION

**Chibuzor Henry Amadi[1], Akaniyene Benard Obot[2], Kufre Monday Udofia[3], Olaoluwa Ayodeji Adegboye[4]**
Department of Electrical and Electronic Engineering,
Imo State University Owerri, Imo State[1], University of Uyo, Akwa Ibom State, Nigeria[2,3,4]
chamadi@imsuoline.edu.ng[1],akaniyeneobot@uniuyo.edu.ng[2], kmudofia@uniuyo.edu.ng[3], mathsgene@yahoo.com[4]

## ABSTRACT

In this study, analytical modelling of the impact of foliage cover on the propagation loss in a smart farming Internet of Things (IoT) application is presented. The Weissberger's exponential decay model and free space path models were used to derive the total propagation loss that was made up of the sum of the path loss due to vegetation cover and path loss due to free space path. The propagation loss losses were related analytically to the foliage cover factor, δ and foliage height, hf above the IoT sensor node antenna, Python program was used to simulate the models using total path length, d=2 km, signal frequency, f=2400 MHz, base station antenna height, h = 30 m, foliage cover factor δ ranging from 0.0 to 0.95 and foliage height, hf ranging from 15 m. The results show that when δ reaches 0.95, the free space path distance, $d_{fsp}$ decreases to 0.1 km (that is 100 m) while the foliage cover distance, $d_f$ increases to 1.9 km (that is 1900 m). At this point, the free space path loss, $L_{fsp\delta}$ decreases to 80.05422483 dB while the foliage cover path loss, $d_f$ increases to 144.4570531 dB giving a total path loss of 224.5112779 dB. This is about 2.1 times the total path loss when there was no foliage cover. In all, the results show that both foliage height, hf and foliage cover distance, hf affect the path loss that can be experienced by wireless signal propagating in smart farming environment with foliage cover. The analytical model presented in this study will enable designers of IoT for smart farm applications to effectively quantify the expected path loss in such applications.

## 1. INTRODUCTION

Nowadays, increasing number of researchers and farmers are interested in adopting smart farming technologies to minimize losses, increase productivity and minimize cost and wastages in the farming process [1,2]. Such emerging farming technologies on the use of wireless Internet of Thing (IoT) sensor network where sensor nodes are deployed in the farm environment to acquire and transmit relevant data on timely basis [3,4,5].

Generally, wireless signal experience losses when transmitted over a distance in free space [6,7]. However, when deployed in areas with foliage cover, additional propagation loss is experienced due to the foliage obstruction of the signal path [8,9]. To effectively communicate with remote servers, the sensor node must have adequate signal transmission power that will cater for the signal power losses in the course of the signal processing and propagation [10,11]. Notably, plants are living things and as the plant grows the vegetation cover height increases. This increase in foliage cover height also affects the propagation loss. Accordingly, his works presented analytical models that account for the propagation losses in farm area with emphasis on the effect of foliage cover factor and the variation in the foliage cover height [12,13]. The models are simulated using program written in Python programming language. The study contributed immensely to the body of knowledge regarding the impact of plant growth on the propagation loss and this is essential for researchers and practitioners seeking to employ IoT sensor network in smart farming applications.

## 2 METHODOLOGY

The study used analytical modelling and simulation to evaluate the impact of foliage cover on the propagation loss expected in a smart farming environment. First, the foliage cover factor is modeled. Secondly, the impact of Foliage cover on the path loss for wireless signal in smart farming environment is modeled. Thirdly, the impact of foliage height on the path loss for wireless signal in agricultural environment is modeled. Then the simulation, results and discussions are presented.

### 2.1 Modelling of the Foliage Cover Factor

Consider a precision smart farming environment with foliage height, $h_f$ above the sensor node antenna and base station antenna height, $h_{fsp}$ above the foliage, as shown in Figure 1. Let $G_L$ denote the base station or gateway location, $SN_L$ denotes the sensor node location. Let $d_{fsp}$ denotes the distance between $G_L$ and $SN_L$, covered by free space, and let $d_f$ denote the distance between $G_L$ and $SN_L$, covered by foliage.

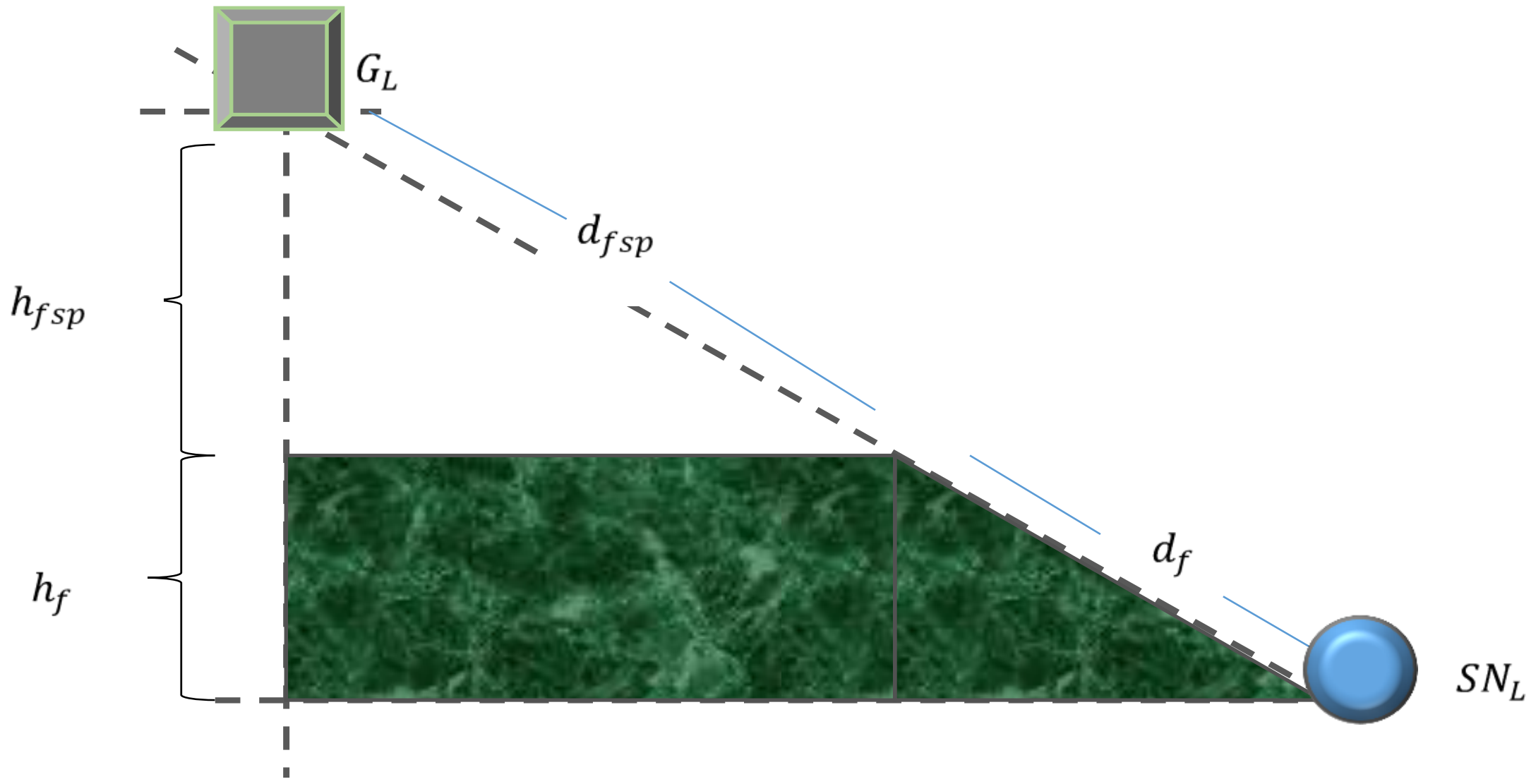

**Figure 1: The precision agricultural space.**

Now the foliage cover factor which is the degree of foliage cover along the transmission path is denoted as $\delta$ defined as;

$$\delta = \frac{d_f}{d} \tag{1}$$

where d is the total path length between the sensor and the base station or gateway. The free space path length denoted as $d_{fsp}$ is given as;

$$d_f = \delta(d) \tag{2}$$

$$d_{fsp} = d - d_f = d(1 - \delta) \tag{3}$$

Consider the communication path which consists of distance covered with free space and distance covered with foliage, as shown in Figure 1. At 100% foliage cover, $d_{fmax} = 400m \Longrightarrow \delta_{max} = \frac{d_{fmax}}{d} = \frac{400}{d}$. The selection of $d_f$ is valid for the constraint$\delta_{min} \leq \delta \leq \delta_{max}$, where $\delta_{min} = 0.01$, and $\delta_{max} = 1$. The maximum distance covered by foliage $\delta_{max}$ can be expressed as;

$$\alpha_H \pm \sigma \cdot \alpha_H \leq \delta_{max} \quad (4)$$

where, $\alpha_H$ denotes a certain number selected as $\delta$, $\sigma$ denotes the percentage increment or decrement on $\alpha_H$ as;

$$\alpha_H(1+\sigma) \leq \delta_{max} \quad (5)$$

Supposed $\alpha_H$ is increased by 50%, then,

$$1.5\alpha_H \leq \delta_{max} \Longrightarrow \alpha_H \leq 0.66\delta_{max} \quad (6)$$

Supposed $\alpha_L$ is decreased by 50%, then,

$$\alpha_L - \sigma \cdot \alpha_L \geq \delta_{min} \quad (7)$$

where, $\alpha_L$ denotes a certain number selected as $\delta$ as;

$$\alpha_L(1-\sigma) \geq \delta_{min} \quad (8)$$

$$0.5\alpha_L \geq \delta_{min} \Longrightarrow \alpha_L \geq 2\delta_{min} \quad (9)$$

Hence, in this paper, the range of values of $\delta$ used for the simulation is as expressed in Equation 9.

## 2.2 Modeling the impact of Foliage Cover on the Path Loss for Wireless Signal in Smart Farming Environment

It is essential to consider the effect of foliage on transmission in a smart farming area. Hence, this research applies the Weissberger's exponential decay model to obtain the propagation loss in this scenario. The loss as a result of foliage can be computed based on Weissberger model as follows [14,15];

$$L_{f\delta} = \begin{cases} 1.33f^{0.284} \cdot [\delta(d)]^{0.588}, & \text{if } 14 < [\delta(d)] \leq 400 \\ 0.45f^{0.284} \cdot [\delta(d)] & \text{if } 0 < [\delta(d)] \leq 14 \end{cases} \quad (10)$$

where, f denotes transmission frequency, $\delta(d)$ denotes depth of foliage. In addition, the free space path loss model (in decibels) , denoted as $L_{fsp}$ is defined as [16,17];

$$L_{fsp} = 32.5 + 20\log_{10}(d) + 20\log(f) \quad (11)$$

Then, $L_{fsp\delta}$, which is the free space path loss model (in decib els) taking into consideration the foliage depth factor, $\delta$ is defined as;

$$L_{fsp\delta} = 20\log_{10}\big(d(1-\delta)\big) + 20\log_{10}(f) - 32.5 \quad (12)$$

Where;

$$d = d_{fsp} + d_f = d(1-\delta) + d_f = d(1-\delta) + \delta(d) \quad (13)$$

Hence, the total path loss, $L_{path}$ in the smart farming area can be computed according to as [18,19,20];

$$L_{path} = L_{f\delta} + L_{fsp} \quad (14)$$

$$L_{path} = 20\log_{10}\big(d(1-\delta)\big) + 20\log_{10}(f) - 32.5 + \begin{cases} 1.33f^{0.284} \cdot [\delta(d)]^{0.588}, & \text{if } 14 < [\delta(d)] \leq 400 \\ 0.45f^{0.284} \cdot [\delta(d)] & \text{if } 0 < [\delta(d)] \leq 14 \end{cases} \quad (15)$$

With this analytical model in Equation 14 or Equation 15, by varying the value of $\delta$, it is possible to determine the impact of foliage cover on the total path loss in the smart farm IoT applications.

## 2.3 Modeling the impact of Foliage Height on the Path Loss for Wireless Signal in Agricultural Environment

Consider the diagram in Figure 1; let h denote the base station antenna height above the sensor antenna height, and d denoted the total distance between the sensor and the base station antenna, then;

$$d = d_f + d_{fsp} \quad (16)$$

$$h = h_f + h_{fsp} \quad (17)$$

where $h_f$ denote the foliage height above the sensor antenna and $h_{fsp}$ denote the base station antenna height above the foliage. Then,

$$h_{fsp} = h - h_f \quad (18)$$

By similar triangle;

$$\frac{h_f}{d_f} = \frac{h_{fsp}}{d_{fsp}} = \frac{h}{d} \quad (19)$$

If the distance, d is contact and the foliage height, $h_f$ above the sensor antenna increases or decreases (that is $h_f$varies), then the $d_f$ and $d_{fsp}$ are given in terms of $h_f$ and d as;

$$d_f = \left(\frac{d}{h}\right) h_f \quad (20)$$

$$d_{fsp} = \left(\frac{d_f}{h_f}\right) h_{fsp} = \left(\frac{d_f}{h_f}\right) (h - h_f) \quad (21)$$

$$d_{fsp} = \left(\frac{d}{h}\right) h_{fsp} = \left(\frac{d}{h}\right) (h - h_f) \quad (22)$$

Now, from Equation 1 and Equation 20, the following expressions apply for δ;

$$\delta = \frac{d_f}{d} = \left(\frac{d}{h}\right) \left(\frac{h_f}{d}\right) = \left(\frac{h_f}{h}\right) \quad (23)$$

$$\delta = \frac{d_f}{d} = \left(\frac{d}{h}\right) \left(\frac{h_f}{d}\right) = \left(\frac{h_f}{h}\right) \quad (23)$$

Hence, the base station antenna height and the foliage height above the sensor node affect the foliage cover factor and hence the path loss experienced by the wireless signal propagating in an agricultural environment.

$$L_{path} = 20\log_{10}\big(d(1-\delta)\big) + 20\log_{10}(f) - 32.5 + \begin{cases} 1.33f^{0.284} \cdot \left[\left(\frac{h_f}{h}\right)(d)\right]^{0.588}, & \text{if } 14 < \left[\left(\frac{h_f}{h}\right)(d)\right] \leq 400 \\ 0.45f^{0.284} \cdot \left[\left(\frac{h_f}{h}\right)(d)\right] & \text{if } 0 < \left(\frac{h_f}{h}\right) \leq 14 \end{cases} \quad (23)$$

## 3. RESULTS AND DISCUSSIONS

In order to evaluate the characteristics of various path losses against the degree of foliage, the key parameter used are set as follows: d = 2 km, f = 2400 MHz, $\delta = 0.0 - 0.95$. The simulation was done in Python 3 and the results obtained are presented and discussed. The graph of foliage cover distance and free space path distance versus foliage cover factor, δ is presented in Figure 2 and it shows that the foliage cover distance is directly proportional to the foliage cover factor, δ whereas the free space path distance is inversely proportional to the foliage cover factor, δ. The foliage cover path loss, free space path loss and total path loss (db) versus foliage cover factor, **δ** is presented in Figure 3. The results show that at foliage cover factor, δ of zero, the free space distance is maximum at 2 km which gave maximum free space path loss of 106.0748247 dB. At this point, the δ =0 and hence the foliage path loss is 0 dB giving a total path loss of 106.0748247 dB. On the other hand, as foliage cover factor, δ increases, the foliage path loss and total path loss increase also while the free space path loss decreases. Specifically, when δ reaches 0.95, the free space path distance, $d_{fsp}$ decreases to 0.1 km (that is 100 m) while the foliage cover distance, $d_f$ increases to 1.9 km (that is 1900 m). At this point, the free space path loss, $L_{fsp\delta}$ decreases to 80.05422483 dB while the foliage cover path loss, $d_f$ increases to 144.4570531 dB giving a total path loss of 224.5112779 dB. This is about 2.1 times the total path loss when there was no foliage cover. This significant increase in the path loss will amount to higher transmitter power requirement with the attendant high energy consumption which cuts down the battery lifespan for the sensor nodes.

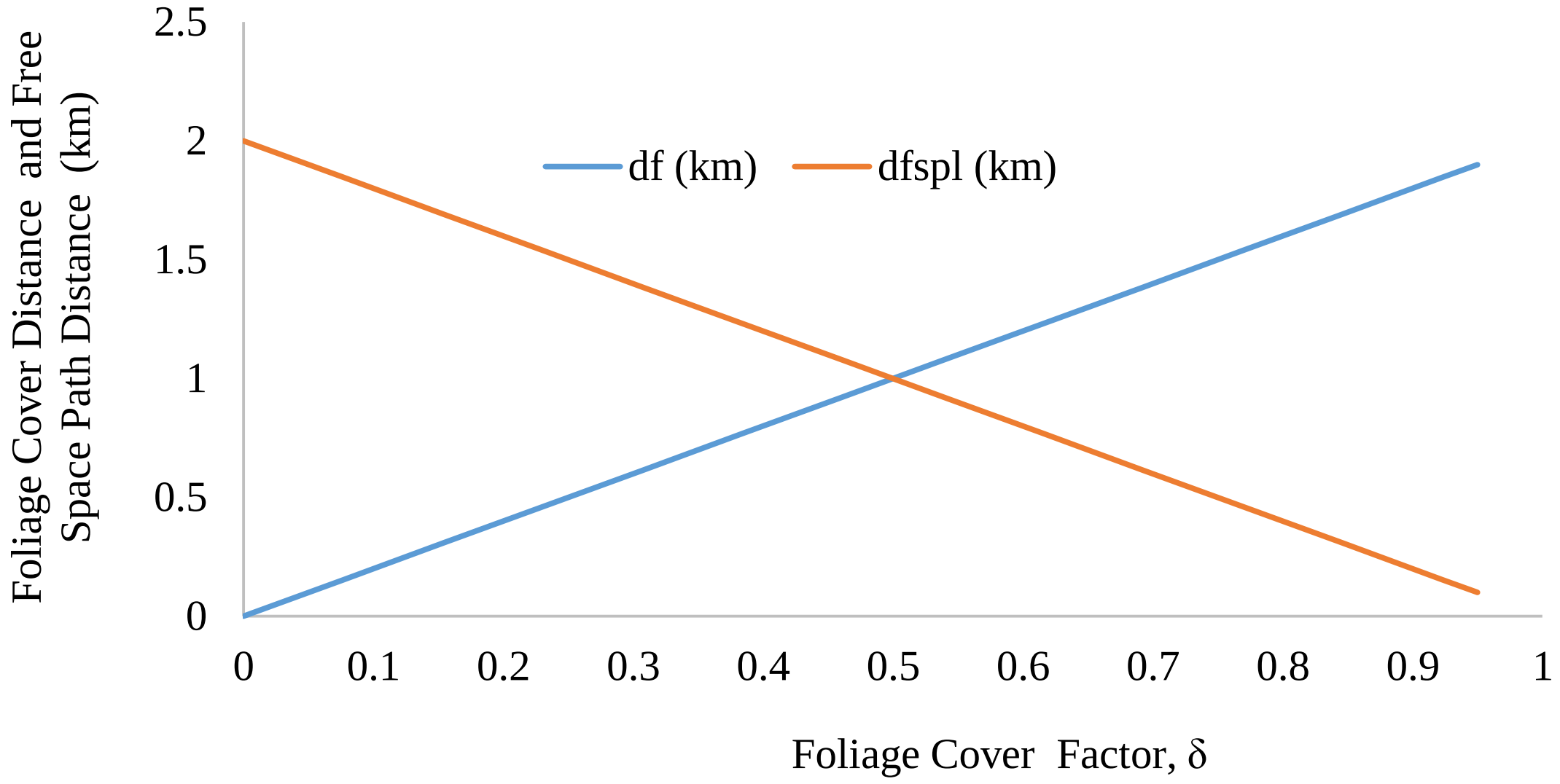

**Figure 2: The graph of foliage cover distance and free space path distance versus foliage cover factor, δ**

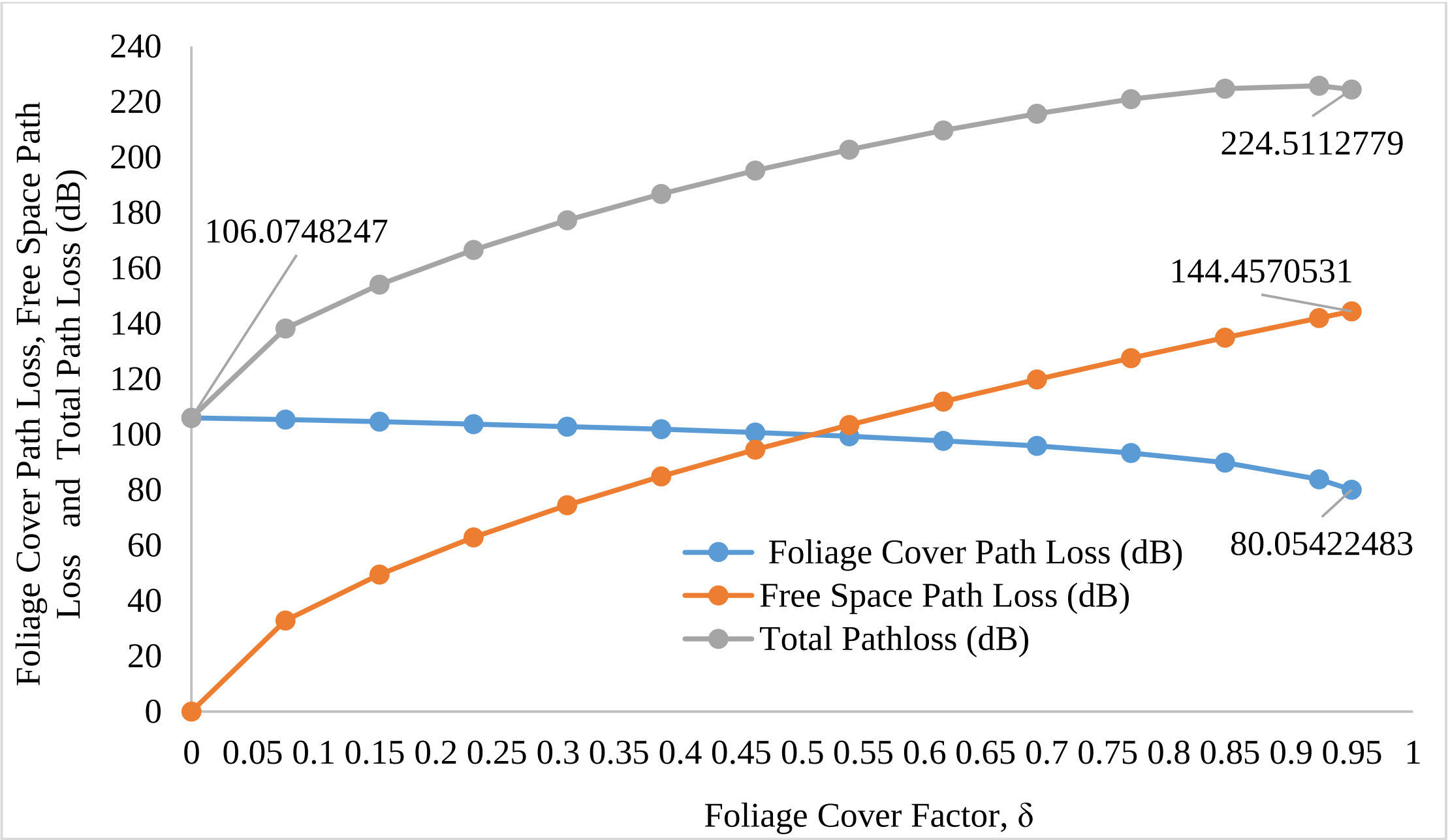

**Figure 3: The foliage cover path loss, free space path loss and total path loss (dB) versus foliage cover factor, δ**

Similarly, the foliage cover path loss, free space path loss and total path loss (db) versus foliage height, hf, is presented in Figure 4, where the total base station antenna height is 30 m. The results show that at foliage height, hf of zero, the free space distance is maximum at 2 km which gave maximum free space path loss of 106.0748247 dB. At this point, hf =0 and hence $\delta = 0$, then the foliage path loss is 0 dB giving a total path loss of 106.0748247dB. On the other hand, as foliage height, hf increases, the foliage path loss and total path loss increase also while the free space path loss decreases. Specifically, when foliage height, hf reaches 15 m which is half of the base station antenna height, h, the foliage cover factor, $\delta$ reaches 0.5, the free space path distance, $d_{fsp}$ decreases to 1 km (that is 1000 m) while the foliage cover distance, $d_f$ increases to 1.0 km (that is 1000 m). At this point, the free space path loss, $L_{fsp\delta}$ decreases to 80.05422483 dB while the foliage cover path loss, $d_f$ increases to 99.04479 dB giving a total path loss of 199.0990144 dB. This is about 1.876967649 times the total path loss when there was no foliage cover. This significant increase in the path loss will amount to higher transmitter power requirement with the attendant high energy consumption which cuts down the battery lifespan for the sensor nodes.

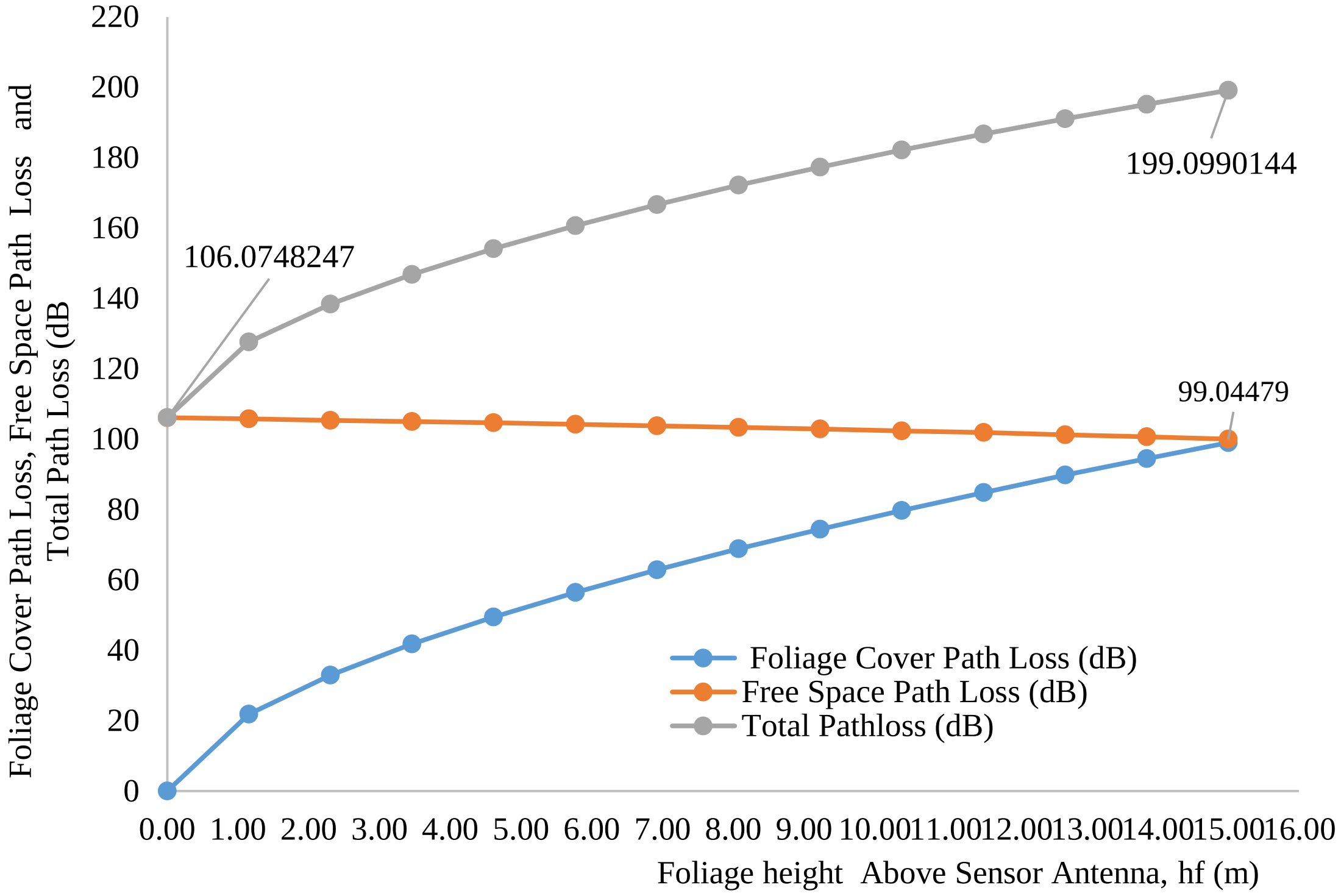

**Figure 4: The foliage cover path loss, free space path loss and total path loss (dB) versus foliage height above sensor antenna, hf (m)**

## 4. CONCLUSION

The impact of vegetation cover on the path loss for wireless sensor node deployed in smart farming area is studied. The relevant analytical expressions are presented for estimating the propagation loss with and without vegetation cover. Simulations were conducted using program written in Python. In all, the results show that foliage height, hf and foliage cover distance, hf affect the path loss that can be witnessed by wireless signal propagating in smart farming environment with foliage cover. The analytical model presented in this study will enable designers of IoT for smart farm applications to effectively quantify the expected path loss in such applications.

## 5. RECOMMENDATIONS

It is also recommended that both the digital wireless network planners and industry should adopt the Weissberger's model for the computation of foliage (vegetation) path loss; this model has a laudable impact on energy consumption. This study enables them to ensure that the selected parameters will not violate the specified error performance when subjected to variations in the available designs.